\PassOptionsToPackage{pdfpagelabels=false}{hyperref}
\documentclass[fleqn,usenatbib,usedcolumn]{mnras}
\usepackage[british]{babel}             

\usepackage{newtxmath,newtxtext}
\DeclareSymbolFont{operators}{OT1}{ntxtlf}{m}{n}
\SetSymbolFont{operators}{bold}{OT1}{ntxtlf}{b}{n}

\usepackage[T1]{fontenc}

\usepackage{graphicx}    
\usepackage{amsmath}    
\usepackage{amssymb}    
\usepackage{xspace}
\usepackage{xstring}
\usepackage{pifont}
\usepackage{pgfkeys}

\newcommand{\rv}{\ensuremath{v_{\textrm{rad}}}\xspace}
\newcommand{\ofe}{\ensuremath{[\textrm{O}/\textrm{Fe}]}\xspace}

\newcommand{\cfe}{\ensuremath{[\textrm{C}/\textrm{Fe}]}\xspace}
\newcommand{\nfe}{\ensuremath{[\textrm{N}/\textrm{Fe}]}\xspace}
\newcommand{\feh}{\ensuremath{[\textrm{Fe}/\textrm{H}]}\xspace}

\newcommand{\ebv}{\ensuremath{\textrm{E}(\textrm{B}-\textrm{V})}\xspace}
\newcommand{\mM}{\ensuremath{\textrm{m}-\textrm{M}}\xspace}

\newcommand{\teff}{\ensuremath{\textrm{T}_\textrm{eff}}\xspace}
\newcommand{\logg}{\ensuremath{\log \textrm{g}}\xspace}
\newcommand{\kms}{\ensuremath{\textrm{km}\,\textrm{s}^{-1}}\xspace}
\newcommand{\eso}{ESO280-SC06\xspace} %
\newcommand{\masyr}{\ensuremath{\textrm{mas}\,\textrm{yr}^{-1}}\xspace}
\newcommand{\bprp}{\ensuremath{\textrm{G}_\textrm{\textsc{bp}}-\textrm{G}_\textrm{\textsc{rp}}}\xspace}
\newcommand{\gaia}{{\it Gaia}\xspace}
\newcommand{\ewcat}{\ensuremath{\textrm{EW}_{_\textrm{CaT}}}\xspace}

\newcommand{\cnstarfull}{6719598900092253184\xspace}
\newcommand{\cnstar}{\cnstarfull}
\newcommand{\response}[1]{{{#1}}}
\defcitealias{Simpson:2018cm}{S18}

\title[A NEMP star in a globular cluster]{A Nitrogen-Enhanced Metal-Poor star discovered in the globular cluster \eso}
\author[J. D. Simpson \& S. L. Martell]{
\parbox{\textwidth}{\raggedright
Jeffrey~D.~Simpson$^{1}$\thanks{Email: \texttt{jeffrey.simpson@unsw.edu.au}}
and Sarah~L.~Martell$^{1,2}$
}
\\
$^{1}$School of Physics, UNSW, Sydney, NSW 2052, Australia\\
$^{2}$Centre of Excellence for Astrophysics in Three Dimensions (ASTRO-3D), Australia\\
}

\date{Accepted 2019 September 13. Received 2019 September 13; in original form 2019 August 28}

\pubyear{2019}

\begin{document}
\label{firstpage}
\pagerange{\pageref{firstpage}--\pageref{lastpage}}
\maketitle

\begin{abstract}
We report the discovery of the only very nitrogen-enhanced metal-poor star known in a Galactic globular cluster. This star, in the very metal-poor cluster \eso, has $\nfe>+2.5$, while the other stars in the cluster show no obvious enhancement in nitrogen. Around 80 NEMP stars are known in the field, and their abundance patterns are believed to reflect mass transfer from a binary companion in the \response{asymptotic giant branch} phase. The dense environment of globular clusters is detrimental to the long-term survival of binary systems, resulting in a low observed binary fraction among red giants and the near absence of NEMP stars. We also identify the first known horizontal branch members of \eso, which allow for a much better constraint on its distance. We calculate an updated orbit for the cluster based on our revised distance of $20.6 \pm 0.5~$kpc, and find no significant change to its orbital properties.
\end{abstract}

\begin{keywords}
globular clusters: individual: \eso
\end{keywords}



\section{Introduction}\label{sec:intro}

Galactic globular clusters have been the subject of intense research over the last two decades \citep[see the reviews of][and references therein]{Gratton2004,Gratton2012}. The main driver is the star-to-star abundance dispersions that are observed for light elements in nearly\footnote{A notable exception is Ruprecht 106 \citep{Villanova:2013fk,Dotter2018}.} all ancient globular clusters. Spectroscopic, astrometric, and photometric observations have been undertaken with the aim of confirming, constraining, and ruling out the various proposed formation mechanisms of these multiple stellar populations \citep[for an overview of the various proposals see the reviews of][and references therein]{Charbonnel2016,Bastian:2018kj,Forbes2018}. As the `typical' clusters have not shown a clear path forward, there is now much interest in exploring the edges of the parameter space: the least massive clusters \citep[e.g.,][]{Mucciarelli:2016fd,Simpson:2017iv}, the `youngest' clusters \citep[e.g.,][]{Valcheva:2014ix,Hollyhead:2017hg}, and young massive clusters in other galaxies that are thought to be to precursors to the ancient globular clusters of the Milky Way \citep[e.g.,][but see discussion in \citealt{Renaud2018,Renaud2019} on whether these objects are truly analogues to present-day globular clusters]{CabreraZiri:2016bm}. It is in this context that we have been investigating the very metal-poor globular cluster \eso.

Prior to \citet[][hereafter \citetalias{Simpson:2018cm}]{Simpson:2018cm}, \eso had been the subject of only two papers that discussed the cluster in any detail: \citet{Ortolani:2000ut, Bonatto:2007be}. In \citetalias{Simpson:2018cm} the knowledge of the cluster was greatly expanded: using new photometry and spectroscopy we found it to be very metal poor ($\feh=-2.47\substack{+0.06 \\ -0.12}$), with a sparsely-populated giant branch, and appearing to lack a horizontal branch.

Subsequent to \citetalias{Simpson:2018cm}, there have been two major developments: firstly, we have undertaken additional spectroscopic observations of \eso; and secondly, the second data release of \gaia \citep{GaiaCollaboration:2016cu,GaiaCollaboration:2018io} has made it possible to use proper motions and improved photometry to refine the cluster membership. As a consequence, one star (\gaia DR2 \texttt{source\_id}=\cnstarfull) previously determined by \citetalias{Simpson:2018cm} to be a field star, should now be re-classified as a member of the cluster. This star is of great interest because it has very strong cyanogen (CN) spectral features for a star of its metallicity and evolutionary stage. In \citetalias{Simpson:2018cm} it was estimated that the star had $\nfe\sim3$ if it were a member of the cluster.

Moderately nitrogen-enhanced ($0<\nfe<2$) stars are well-known in globular clusters and dwarf galaxies \citep[e.g.,][]{Simpson:2012kc,Simpson:2013ee,Simpson:2017be,Marino:2012eg,Carretta2014,Roederer2015,Lardo:2015hi,Gerber2018}. Nitrogen-enhanced stars are classified as belonging to the ``second population''\footnote{The stars in clusters with abundance patterns like the Galactic halo (e.g., low nitrogen and high oxygen) are commonly referred in globular cluster research to as `primordial' or `first population' stars. The stars with enhanced abundances of nitrogen and depleted oxygen are `enriched' or `second population' stars.} of stars in a globular cluster. Strong CN features can be observed at relatively low spectral resolution, and enhanced nitrogen has been used to chemically tag a small fraction of halo and bulge stars as having initially formed in globular clusters \citep[e.g.,][]{Martell:2010is,FernandezTrincado:2016db,Fernandez-Trincado2017,Schiavon:2017dg,Tang2018}.

The moderate nitrogen enhancements in globular cluster stars are produced by a combination of primordial (`second population' enhancement) and evolutionary \cite[`extra mixing'; e.g.,][]{Denissenkov2003} processes. But there is a class of metal-poor ($\feh<-2$) field stars with even higher levels of nitrogen enhancement --- nitrogen-enhanced metal-poor (NEMP) stars, defined by \citet{Johnson2007} as having $\nfe>+0.5$ and $[\mathrm{N}/\mathrm{C}]>0.5$. The high \nfe abundances in these stars are believed to be the result of mass transfer from an intermediate-mass \response{asymptotic giant branch (AGB)} companion \citep[e.g.,][]{Pols:2012bm}.

This `NEMP' definition encompasses some of the nitrogen-enhanced stars in metal-poor globular clusters, which implies that such field stars could likely be explained as `second population' stars lost from globular clusters. However, primordial enhancement followed by mixing cannot increase the nitrogen abundance enough to produce stars with $\nfe>+2$. NEMP stars with such high \nfe abundances are rare in the field, with $\sim15$ known \citep[per the SAGA database;][]{Suda2008}, compared to carbon-enhanced metal-poor stars, of which over 300 have been discovered \citep{Yoon2016}. If the \nfe estimate from \citetalias{Simpson:2018cm} of \cnstarfull can be confirmed, it would be the first such very nitrogen-enhanced metal-poor star known in a globular cluster.

In Section \ref{sec:spec_obs} we describe the spectroscopic observations, data reduction, and analysis; in Section \ref{sec:cluster_members}, we define the criteria for cluster membership; in Section \ref{sec:bulk} we determine various bulk properties of the cluster and its extra-tidal stars; in Section \ref{sec:cn-star} we estimate carbon and nitrogen abundances for bright giants in \eso, and in section \ref{sec:discuss} we discuss the NEMP star, the implications of its apparent uniqueness in the Galaxy, and the further work needed to fully understand its origins.

\section{Observational data and parameter determination}\label{sec:spec_obs}

\begin{figure*}
    \includegraphics[width=\textwidth]{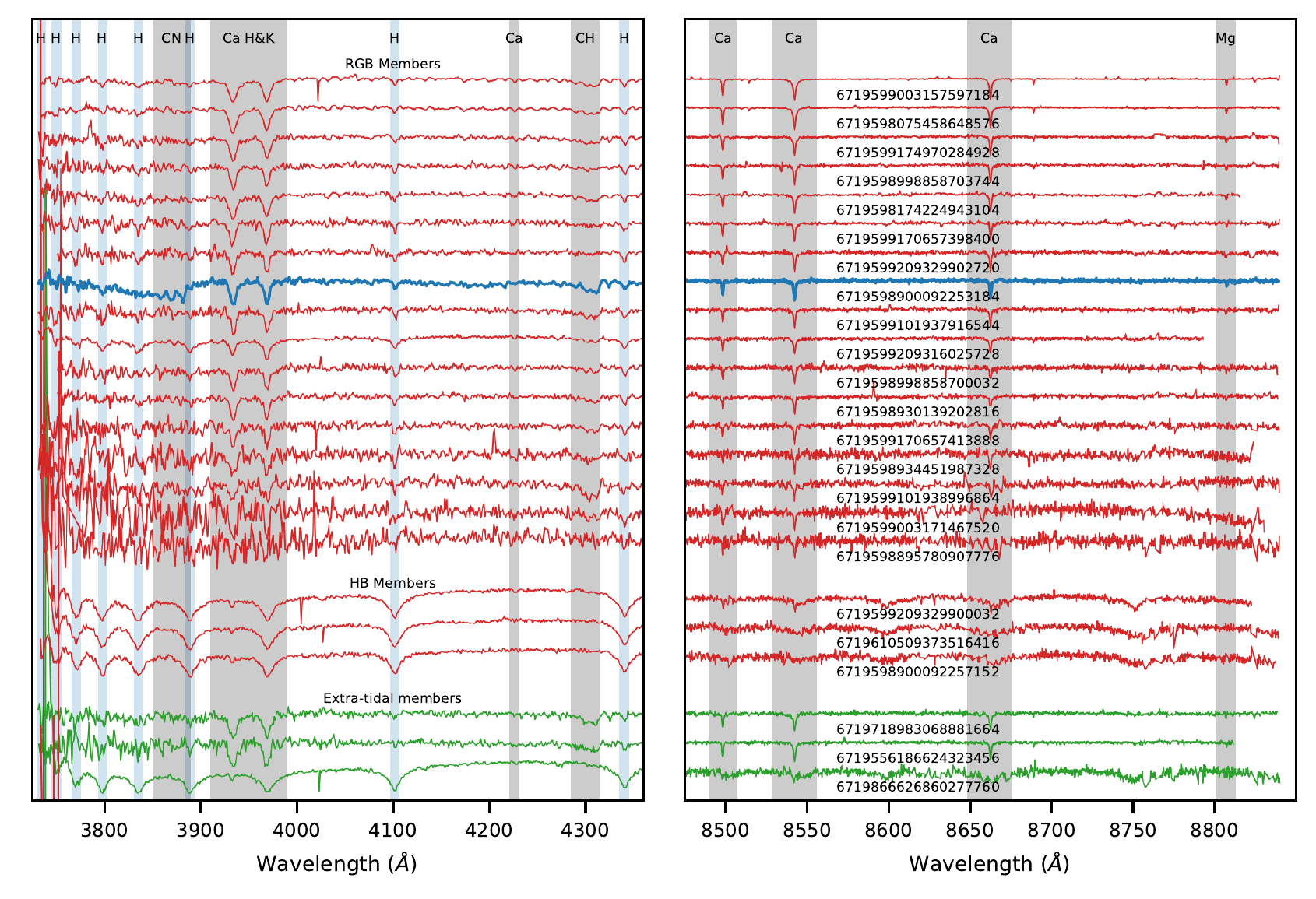}
    \caption{Reduced, continuum-normalized spectra for the identified members of \eso. The left panel is the wavelength region of interest of the blue spectrum, containing the CN and CH molecular bands. The right panel shows the red spectrum and the CaT lines. The spectra are divided between the RGB members, the HB members, and the extra-tidal stars. Within in each section the stars are ordered from top-to-bottom by increasing G magnitude. One of the RGB members (\cnstarfull) is highlighted in blue with a thicker line: this is the very CN-strong star.}
    \label{fig:spec_stacked}
\end{figure*}

To supplement the observations undertaken for \citetalias{Simpson:2018cm}, \eso was observed on the nights of 2018 June 5 and 6 with the 3.9-metre Anglo-Australian Telescope and its AAOmega spectrograph \citep{Sharp:2006bd}, with the 392-fibre Two Degree Field fibre positioner (2dF) top-end \citep{Lewis:2002eg}.

AAOmega is a moderate resolution, dual-beam spectrograph. As in our previous cluster work \citep{Simpson:2017be,Simpson:2017iv,Simpson:2018cm}, the following gratings were used at their standard blaze angles: the blue 580V grating ($R\sim1200$; 3700--5800~\AA) and red 1700D grating ($R\sim10000$; 8340--8840~\AA). The 580V grating provides low-resolution coverage of the calcium H \& K lines and spectral regions dominated by CN and CH molecular features in cool giants. The 1700D grating was specifically designed to observe the near-infrared calcium triplet ($\sim8500$~\AA) at high resolution for precise radial velocity measurements.

A total of 687 stars were observed across two field configurations. These observations targeted the stars identified using \gaia DR2 as being possible cluster members based upon their proper motions, with a particular focus on possible HB stars, as this region of the CMD had not been targeted previously. For each field there were 25 sky fibres, and flat lamp (40~s) and arc lamp (60~s; Fe+Ar, Cu+Ar, Cu+He, Cu+Ne) exposures were acquired. The science observations on both nights were six 1800-sec exposures. The observations on 2018 June 5 had seeing between 1.5--1.9~arcsec, and on June 6 1.9--4.0~arcsec. The raw images were reduced to 1D spectra using the AAO's \textsc{2dfdr} data reduction software \citep[][v6.46]{AAOSoftwareTeam:2015vz} with the default configuration appropriate for each grating. This performs all of the standard steps for reducing multi-fibre data. Examples of the reduced spectra can be seen in Figure \ref{fig:spec_stacked}.

From all of our spectroscopic observations of \eso\ \citep[this work and][]{Simpson:2018cm}, we observed 1669 stars within about 1~deg of \eso. All the reduced spectra were processed with a pipeline developed for Simpson (2020, in prep) that uses the near-infrared calcium triplet (CaT) lines at 8498.03, 8542.09 and 8662.14 \AA\ \citep{Edlen:1956ud} to measure the equivalent widths of the CaT lines (\ewcat) and radial velocities of the stars. Briefly, a simple template spectrum is constructed from three pseudo-Voigt functions (the sum of a Gaussian and a Lorentzian function). These pseudo-Voigt functions are simultaneously fit to the CaT lines to find their equivalent widths. This fitted template spectrum is then cross-correlated with the observed spectrum to find the radial velocity of the star. This process is repeated 100 times with random noise inserted into the spectrum proportional to the variance in each pixel. This method works well for giant branch stars, but not for horizontal branch stars. In HB stars, the CaT lines are very weak, and the region is dominated by the hydrogen Paschen lines. As such we used a modification of the above method for HB stars, but using the low resolution blue spectrum, and the Balmer lines. For the stars identified as horizontal branch stars, their \ewcat is arbitrarily set to zero in the subsequent analysis.

The stars were positionally cross-matched with the \gaia DR2 catalogue. In the \gaia DR2 data-set stars within crowded regions can suffer from source confusion which could affect the quality of their photometry \citep{Evans:2018cj} and/or astrometric solution \citep{Lindegren2018}. We define `good' \gaia photometry as those stars with
\begin{equation}\label{eq:good_photom}
	1.0 + 0.015(\bprp)^2<\texttt{EF}<1.3 + 0.06(\bprp)^2,
\end{equation}
where \texttt{EF} is the \texttt{phot\_bp\_rp\_excess\_factor} \citep[these limits are taken from][]{Babusiaux:2018di}. Of the observed stars, 94 per cent (1580/1683) meet this criterion. We define stars with `good' astrometry are those for which
\begin{equation}\label{eq:good_astrom}
	\texttt{RUWE}<1.4,
\end{equation}
where \texttt{RUWE} is the Renormalised Unit Weight Error, defined in \citet{Lindegren2018}, who also recommend the $1.4$ limit. Of the observed stars, 96 per cent (1608/1683) meet this criterion. A total 91 per cent (1527/1683) meet both criteria.

\section{Cluster membership}\label{sec:cluster_members}
\begin{table*}
\centering
\caption{Observed stellar parameters for the stars shown in Figure \ref{fig:spec_stacked}.}
\label{table:star_params}
\begin{tabular}{rrrrrrrrrr}
\hline

\response{\gaia DR2} \texttt{source\_id} & G & \bprp & \rv & $e(\rv)$ & \ewcat & $e(\ewcat)$ & \feh & Radial distance & Membership \\
 & &  & (\kms) & (\kms) & (\AA) & (\AA) &  & (arcmin) &  \\
\hline
6719599003157597184&$14.13$&$1.56$&$93.77$&$0.07$&$4.62$&$0.03$&$-2.25$& 0.13& RGB Member\\
6719598075458648576&$14.61$&$1.51$&$95.54$&$0.07$&$4.31$&$0.02$&$-2.27$& 2.47& RGB Member\\
6719599174970284928&$15.83$&$1.28$&$93.96$&$0.39$&$3.08$&$0.07$&$-2.49$& 0.84& RGB Member\\
6719598998858703744&$15.95$&$1.25$&$90.90$&$0.49$&$2.97$&$0.11$&$-2.51$& 0.51& RGB Member\\
6719598174224943104&$16.03$&$1.35$&$94.12$&$0.30$&$2.98$&$0.05$&$-2.49$& 1.67& RGB Member\\
6719599170657398400&$16.36$&$1.15$&$96.65$&$0.42$&$2.15$&$0.08$&$-2.75$& 1.19& RGB Member\\
6719599209329902720&$16.45$&$1.15$&$94.95$&$0.46$&$2.77$&$0.12$&$-2.51$& 0.27& RGB Member\\
6719598900092253184&$16.47$&$1.30$&$92.14$&$0.44$&$3.00$&$0.10$&$-2.42$& 1.31& RGB Member\\
6719599101937916544&$16.55$&$1.24$&$95.61$&$0.65$&$2.45$&$0.10$&$-2.61$& 0.36& RGB Member\\
6719599209316025728&$16.69$&$1.12$&$96.94$&$0.35$&$2.56$&$0.09$&$-2.55$& 0.10& RGB Member\\
6719598998858700032&$17.19$&$1.12$&$96.45$&$0.96$&$2.43$&$0.16$&$-2.52$& 0.15& RGB Member\\
6719598930139202816&$17.33$&$1.21$&$93.42$&$0.57$&$3.01$&$0.17$&$-2.28$& 0.95& RGB Member\\
6719599170657413888&$17.68$&$1.17$&$94.34$&$0.90$&$2.37$&$0.15$&$-2.48$& 0.98& RGB Member\\
6719598934451987328&$17.77$&$1.20$&$92.23$&$1.32$&$1.97$&$0.17$&$-2.62$& 1.04& RGB Member\\
6719599101938996864&$18.42$&$1.12$&$97.95$&$2.35$&$1.99$&$0.28$&$-2.53$& 0.37& RGB Member\\
6719599003171467520&$18.46$&$1.11$&$91.00$&$2.02$&$2.21$&$0.26$&$-2.43$& 0.39& RGB Member\\
6719598895780907776&$18.97$&$1.22$&$96.83$&$2.14$&$2.41$&$0.87$&$-2.28$& 1.04& RGB Member\\
6719599209329900032&$17.38$&$0.45$&$83.91$&$4.91$&$0.00$&&& 0.71& HB Member\\
6719610509373516416&$17.82$&$0.22$&$110.43$&$4.63$&$0.00$&&& 2.29& HB Member\\
6719598900092257152&$18.05$&$0.26$&$94.40$&$6.40$&$0.00$&&& 0.36& HB Member\\
6719718983068881664&$15.93$&$1.20$&$118.01$&$0.53$&$3.55$&$0.15$&$-2.31$& 41.59& Extra-tidal member\\
6719556186624323456&$16.22$&$1.22$&$85.76$&$0.43$&$3.27$&$0.12$&$-2.36$& 27.85& Extra-tidal member\\
6719866626860277760&$17.52$&$0.39$&$91.35$&$5.99$&$0.00$&&& 33.81& Extra-tidal member\\
\hline
\end{tabular}
\end{table*}

\begin{figure}
    \includegraphics[width=\columnwidth]{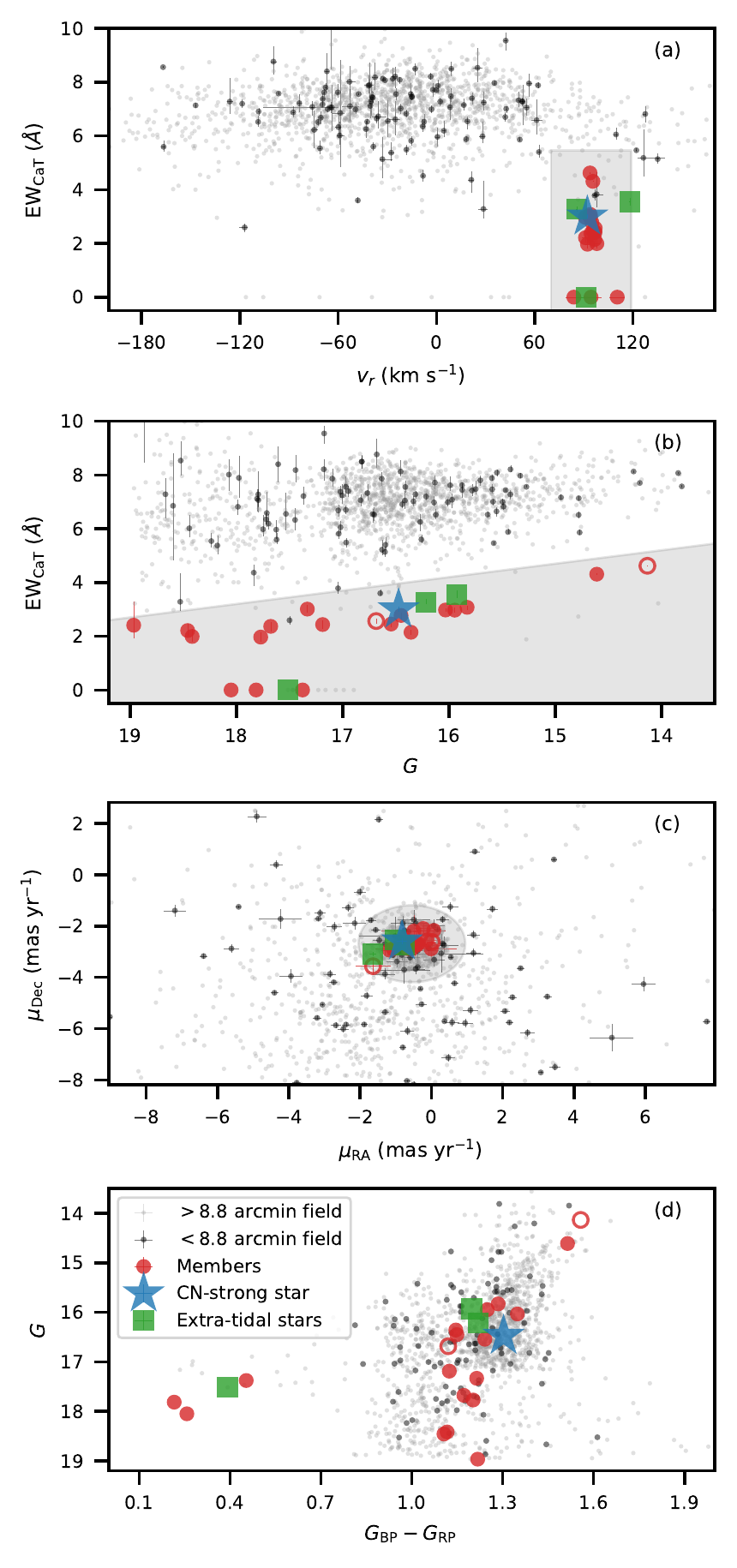}
    \caption{Distribution of the parameters of the spectroscopically observed stars within 1~deg of \eso: (a) radial velocity ($v_r$) versus the equivalent width sum of the three calcium triplet lines (\ewcat); (b) \gaia $G$ magnitude versus the \ewcat; (c) the proper motion distribution; (d) the colour-magnitude diagram from \gaia photometry. Horizontal branch stars are shown with $\ewcat\equiv0$. Unfilled symbols are used for those stars that do not meet the good astrometry or photometry criteria (Equations \ref{eq:good_photom} and \ref{eq:good_astrom}). Members of
    \eso (red circles) were selected as those within the tidal radius (8.8~arcmin) and with $70~\kms<v_r<120~\kms$; proper motion within 1.5~\masyr of $(\mu_\mathrm{RA},\mu_\mathrm{Dec})=(-0.548,-2.688)~\masyr$; and (if not an HB star) $12.1 - 0.5G > \sum{\ewcat}$. We identified three stars outside the tidal radius that meet the selection criteria and these are shown with green squares.}
    \label{fig:member_selection}
\end{figure}

The measured properties of the spectroscopically observed stars are shown in various parameter spaces in Figure \ref{fig:member_selection} \response{and are given in Table \ref{table:star_params}}.  The vast majority of the observed stars have $\ewcat>4$~\AA, and with a large range of velocities, i.e., we have sampled the Milky Way field population along the line-of-sight. As found in \citetalias{Simpson:2018cm}, there is a small group of stars at $\rv\approx95$~\kms and with low \ewcat (shaded region on Figure \ref{fig:member_selection}a). These stars also have correlated \ewcat and apparent magnitude (Figure \ref{fig:member_selection}b), a trend expected for RGB stars from a mono-metallic, spatially co-located population (e.g., a stellar cluster) --- CaT line strength increases as the star increases in luminosity \citep[e.g.,][]{Armandroff:1991ic}.

The shaded regions on Figure \ref{fig:member_selection}a,b,c indicate our \eso membership selection criteria: stars with $70~\kms<v_r<120~\kms$; proper motion within 1.5~\masyr of $(\mu_\mathrm{RA},\mu_\mathrm{Dec})=(-0.548,-2.688)~\masyr$; and (if not an HB star) $12.1 - 0.5G > \sum{\ewcat}$. In addition, we required that the star be within the tidal radius of 8.8~arcmin (Section \ref{sec:distance} and Table \ref{table:cluster_params}). Stars outside of this radius that met the other criteria are classified as possible extra-tidal stars. In total, we have identified 17 RGB members, three HB members, and three possible extra-tidal members.

Figure \ref{fig:member_selection}d shows the colour-magnitude diagram (CMD) in \gaia photometry of the observed stars. Of note are the three horizontal branch members. This solves a mystery from \citetalias{Simpson:2018cm}, where a lack of HB stars was discussed. The presence of a HB has two important results: (i) it is no longer necessary to invoke unusual stellar evolution modes or preferential mass loss from the cluster to explain the lack of HB members; (ii) the distance to the cluster can be refined (Section \ref{sec:distance}), which could adjust our estimate of the metallicity of the cluster (Section \ref{sec:metallicity}).

\section{Bulk properties of the cluster}\label{sec:bulk}
\begin{table}
\caption{Summary of the observational parameters for \eso determined in this work.}
\label{table:cluster_params}
\begin{tabular}{ll}
\hline
$v_r$ & $94.64\pm0.48$\kms\\
$\sigma_r$ & $2.31\pm0.36$\kms\\
$d_\odot$ & $20.6\pm0.5$~kpc\\
$r_c$ & $14.76\pm0.32$~arcsec or \response{$1.48\pm0.05$~pc}\\
$r_t$ & $8.82\pm2.48$~arcmin or $53.1\pm15.0$~pc\\
$\ebv$ & $0.14\pm0.01$\\
$(m-M)_V$ & $17.01\pm0.04$\\
$(m-M)_0$ & $16.57\pm0.05$\\
$\feh$ & $-2.48\pm0.04$\\
\hline
\end{tabular}
\end{table}

In this section we determine various bulk properties of the cluster: distance and size (Section \ref{sec:distance}), metallicity (Section \ref{sec:metallicity}); radial velocity (Section \ref{sec:radial_velocity}), and orbit about the Milky Way (Section \ref{sec:orbit}).

\subsection{Distance and size}\label{sec:distance}
\begin{figure}
    \includegraphics[width=\columnwidth]{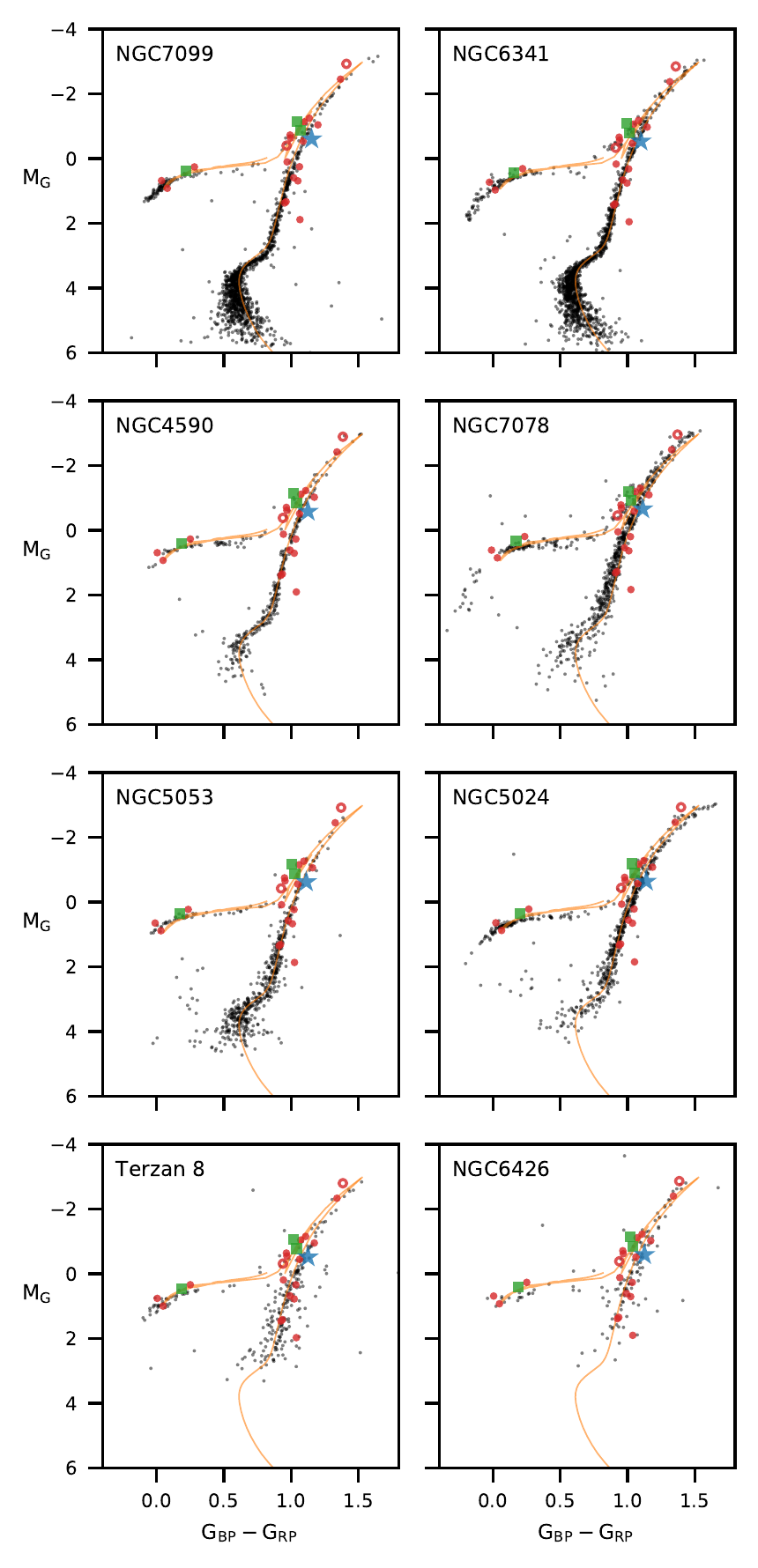}
    \caption{Dereddened-colour absolute-magnitude diagrams of eight metal-poor globular clusters (black dots), created using \gaia DR2 photometry, and with distance moduli and reddening values from the latest compilation by \citet{Harris1997}. For each cluster, by eye, we found the best reddening and distance moduli for \eso members (shown with the same symbols as in Figure \ref{fig:member_selection}). To aid the eye, we also show the MESA Isochrones \& Stellar Tracks \citep[][]{Dotter:2016fa,Choi:2016kf,Paxton:2010jf,Paxton:2013km,Paxton:2015iy} isochrone for $\log(\mathrm{Age})=10.15,~\feh=-2.5$. It was necessary to add $+0.07$ magnitudes to the \bprp colour of the isochrones to have them overlap with the cluster sequences. We found an average distance modulus of $(\mM)_\mathrm{V} = 17.011\pm0.045$ and an average reddening of $\ebv = 0.141\pm0.006$ for \eso from this differential analysis.}
    \label{fig:relative_cmds}
\end{figure}

We are able to refine the distance to \eso using the photometry of \gaia DR2 and the horizontal branch stars identified in this work. Distances to clusters are typically estimated using isochrone fitting, but there is a lack of very metal-poor (i.e., $\feh<-2.3$), $
\upalpha$-enhanced isochrones --- that include the HB --- in \gaia photometry. Instead, we performed a differential analysis with other metal-poor clusters to estimate the distance to \eso.

Eight metal-poor clusters were selected that had well-defined cluster sequences in \gaia photometry. For each cluster, all stars with good photometry \response{and astrometry}, within the tidal radius of the cluster, and within 1~\masyr of the proper motion of the cluster \citep[taken from][]{Vasiliev:2018uf} were selected as `members' from the \gaia DR2 catalogue. Using the distance modulus and reddening for each cluster from \citet{Harris1997}, assuming $A_V = 3.1\times\ebv$, and applying the reddening and extinction corrections from \citet[][namely their equation 1 and table 1]{Babusiaux:2018di}, a dereddened-colour absolute-magnitude diagram was constructed for each cluster (Figure \ref{fig:relative_cmds}). Then by eye, a distance modulus and reddening for \eso relative to each cluster sequence was found.

\textsc{pymc3} \citep{Salvatier2016} was used to fit a Bayesian normal distribution to the distance moduli and reddening values found for \eso with comparison to these eight clusters. This \response{gave} a mean distance modulus for \eso of $(\mM)_\textrm{V} = 17.011\pm0.045$ and an average reddening of $\ebv = 0.141\pm0.006$, i.e., $(\mM)_0 = 16.57\pm0.05$. This is slightly smaller than the value determine by \citetalias{Simpson:2018cm} ($[\mM]_0=16.8\pm0.2$). The reddening is consistent with that estimated from the all-sky reddening maps determined by \citet{Schlegel:1998fw} and \citet{Schlafly:2011iu}, who reported $\ebv= 0.16$ and 0.13, respectively, for the location of \eso.

This distance modulus places \eso at distance from the Sun of $d_\odot = 20.6\pm0.5$~kpc. Transformed into Galactocentric coordinates, $d_\mathrm{GC} = 13.0\pm0.5=[(X,Y,Z)=(11.3\pm0.4,-4.6\pm0.1,-4.5\pm0.1)]$~kpc. The core radius was measured using \textsc{ASteCA} \citep{Perren2015} from the \gaia DR2 star counts. This found  \response{core radius of} $r_c = 14.76\pm0.32$~arcsec and \response{a tidal radius of} $r_t = 8.82\pm2.48$~arcmin. For our measured distance, this equates to a physical size of $r_c=1.5\pm0.0$~pc and $r_t=53.1\pm15.0$~pc.

\subsection{Metallicity}\label{sec:metallicity}
\begin{figure}
    \includegraphics[width=\columnwidth]{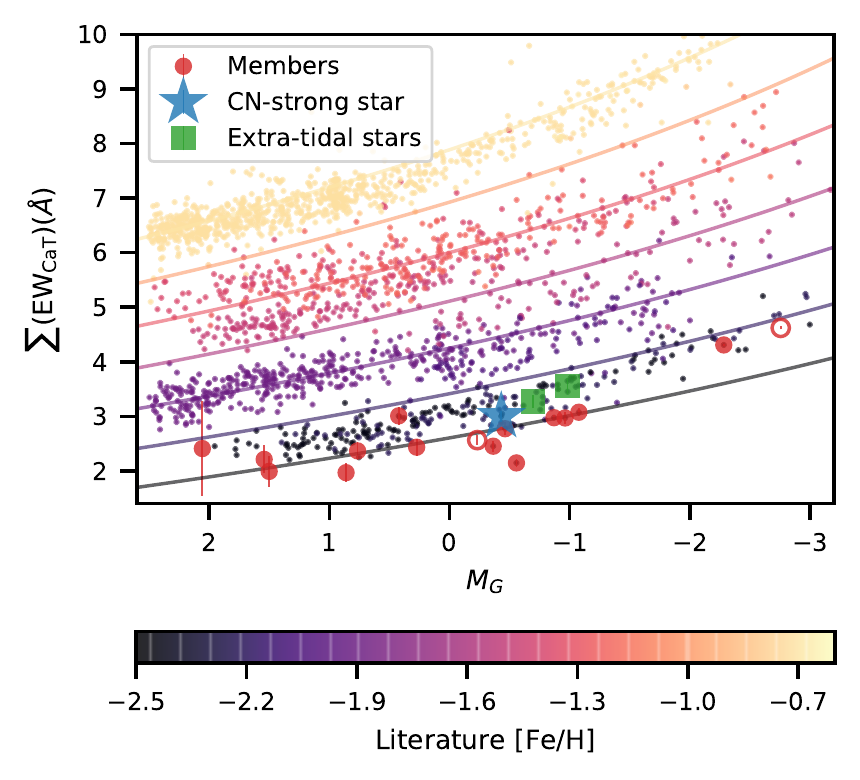}
    \caption{For 2050 RGB stars from 18 clusters, including \eso, we show the \ewcat against their absolute magnitude. For all clusters except \eso the stars are colour-coded by their literature metallicity. The \eso stars are shown using the same symbols as Figure \ref{fig:member_selection}. The solid curves show the empirical relationship found in Equation \ref{eq:metallicity} for $\feh\in[-2.5,-0.7]$ with a step size of 0.3. We find that \eso has lower \ewcat than the other very metal-poor cluster in the sample (NGC7099: $\feh=-2.44$). From the empirical relationship, we find a metallicity of $\feh=-2.48\pm0.04$.}
    \label{fig:metallicity_other_clusters}
\end{figure}

\eso is of great interest as  \citetalias{Simpson:2018cm} found it to be very metal poor: $\feh=-2.47\substack{+0.06 \\ -0.12}$. This is at the apparent floor in the metallicity distribution function for GCs in the Milky Way and local Universe \citep{Kruijssen2019}. To estimate the metallicity of \eso, with the available spectra, we use the empirical CaT line strength method \citep[e.g.,][]{Mauro:2014ik,Carrera:2007kd,Starkenburg2010}.

Here we are able to improve on that analysis used in \citetalias{Simpson:2018cm} with a new empirical CaT-\feh relationship developed for \gaia photometry (Simpson 2020, in prep). Briefly, the Anglo-Australian Telescope archive was searched for observations of globular clusters with AAOmega and the 1700D grating. These spectra were processed in the same way as the spectra of \eso (Section \ref{sec:spec_obs}). A total of 2050 stars from 18 clusters were identified as being cluster members based upon their kinematics, photometry, and \ewcat. These clusters covered a range of metallicities from $\feh=-0.69$ (NGC6624) to $\feh=-2.44$ (NGC7099). Distance moduli, reddening, and metallicities were taken from \citet{Usher2018}. The apparent $G$ magnitudes were converted to absolute magnitudes using \citet{Babusiaux:2018di} as in Section \ref{sec:distance}. Figure \ref{fig:metallicity_other_clusters} shows the results for all of the clusters, with the \ewcat of each star versus its absolute magnitude. \textsc{emcee} \citep{ForemanMackey:2013io} was used to fit the following function to the data
\begin{equation}\label{eq:metallicity}
	\feh=a+bG + c\Sigma(EW) + d[\Sigma(EW)]^2 + eG\Sigma(EW).
\end{equation}
where the best fit found was $a=-3.524_{-0.024}^{+0.022}$, $b=0.108_{-0.007}^{+0.007}$, $c=0.410_{-0.008}^{+0.009}$, $d=-0.007_{-0.001}^{+0.001}$, $e=0.015_{-0.001}^{+0.001}$. The curves on Figure \ref{fig:metallicity_other_clusters} indicate the loci of constant metallicity in this plane, using this function.

Using Equation \ref{eq:metallicity} and the 15 \eso RGB members with good photometry we calculate that \eso has a metallicity of $\feh=-2.48\pm0.04$. This is actually the same value as found in \citetalias{Simpson:2018cm} when using a different set of photometry and calibration. It confirms that \eso is one of the most, if not the most, metal-poor globular clusters in the Galaxy.

\subsection{Radial velocity}\label{sec:radial_velocity}
The mean radial velocity and its dispersion were estimated assuming that the cluster was in virial motion (which considering the sparse nature of the cluster may not be true). A Bayesian normal distribution was fitted to the radial velocities of the 19 RGB stars (the horizontal branch members were excluded as their radial velocities are relatively uncertain), finding $v_r = 94.89\pm0.53~\kms$ and $\sigma_r=2.27\pm0.38~\kms$. These values are slightly larger than that determined in \citetalias{Simpson:2018cm} ($v_r = 92.5^{+2.4}_{-1.6}~\kms$ and $\sigma_r=1.5\pm0.01~\kms$). This is a consequence of that work using a smaller sample of 13 stars.

\subsection{Galactic Orbit}\label{sec:orbit}

\begin{figure*}
	\includegraphics[width=\textwidth]{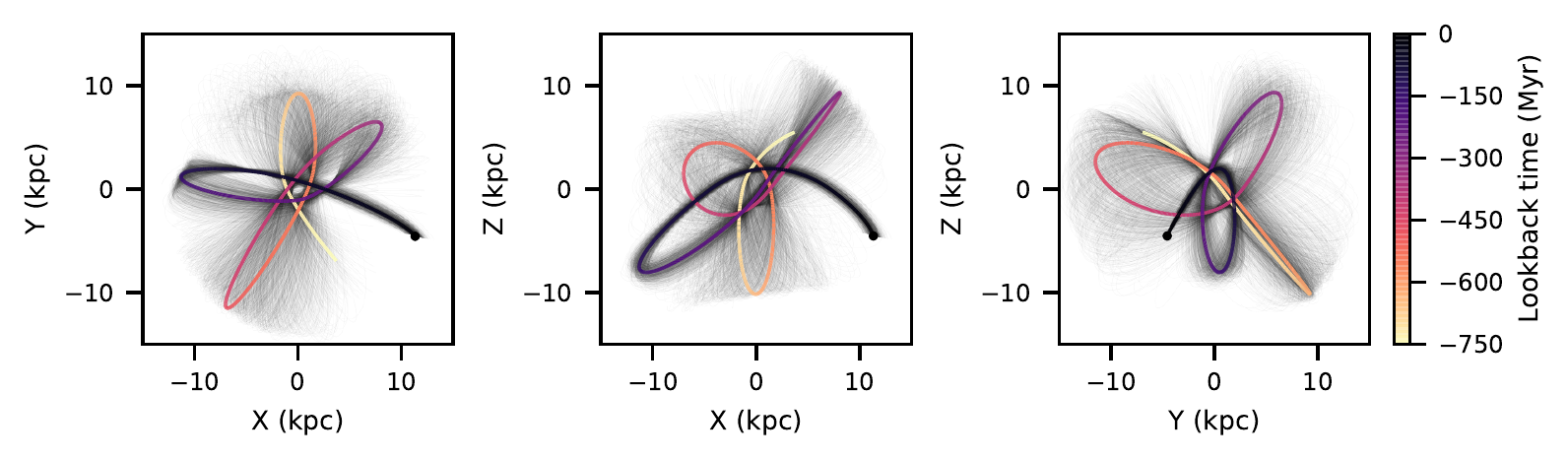}
    \caption{The previous 0.75~Gyr of the orbit of \eso projected into Cartesian space centred on the Galactic Centre. The heavier line is the orbit from the nominal values of the phase space coordinates, colour-coded by lookback time, and the faint black lines show 100 orbits randomly sampling the error distributions of the input parameters. The black dot shows the observed position of \eso.}
    \label{fig:orbit}
\end{figure*}

With the spatial and kinematic information of the cluster now known, we can estimate an orbit for \eso. We used \textsc{gala} \citep[version 1.0;][]{Price-Whelan2017a,Price-Whelan2018b}, with the default potential \texttt{MilkyWayPotential}. This is a simple mass-model for the Milky Way consisting of a spherical nucleus and bulge, a Miyamoto-Nagai disk, and a spherical NFW dark matter halo. The parameters of this model are set to match the circular velocity profile and disk properties of \citet{Bovy:2015gg}.  The Sun's velocity is taken from \citet{Schonrich2012} to be $(U_\odot,V_\odot,W_\odot)=(11.0,248.0,7.25)~\kms$ . Errors in the calculated orbital parameters were estimated by taking 1000 samples of the error distributions and finding the 16th and 84th percentiles of the given results.

We find that the \eso has an eccentric orbit ($e=0.81_{-0.06}^{+0.03}$), with an apocentric distance of $13.36_{-0.98}^{+1.27}~\mathrm{kpc}$ and a pericentric distance of $1.40_{-0.36}^{+0.73}~\mathrm{kpc}$. The cluster is in a moderately prograde orbit ($L_z=-314_{-106}^{+102}~\textrm{kpc}\,\textrm{km}\,\textrm{s}^{-1}$). Figure \ref{fig:orbit} shows the previous 750~Myr of its orbit (coloured line), as well as 100 other possible orbits over the same time interval created by sampling the error distributions (faint black lines). At the present time \eso is just past apocentre, and is sweeping back towards the Galactic centre. Such an orbit is typical of many clusters found in the halo \citep[e.g.,][]{Simpson2019}, i.e., highly eccentric orbits with some time spent in the inner bulge of the Galaxy.

\gaia astrometry has allowed for the orbits of almost every Galactic globular cluster to be calculated. This has lead to several authors \cite[e.g.,][]{Helmi:2018wy,Kruijssen2018,Myeong2019} grouping clusters into in-situ and accreted clusters, and then further grouping accreted clusters by various proposed accretion events. As cautioned by \citet{Piatti2019b}, there is substantial overlap, but also obvious disagreement in their lists of clusters. In the case of \eso, \citet{Massari2019} has associated it with \gaia-Enceladus (a.k.a.\ \textit{The Sausage}). Although we have revised the distance and observed kinematics of the cluster, our orbit is similar enough to the orbit used by \citet{Massari2019} that their classification would hold.

\section{Carbon and nitrogen abundances}\label{sec:cn-star}

In \citetalias{Simpson:2018cm}, we identified $\texttt{source\_id}=\cnstarfull$ as having an anomalously strong CN band in its blue spectrum (see Figure \ref{fig:spec_stacked}), but dismissed it as a field star due to the extremely large nitrogen abundance that was required to explain the CN-band strength (\citetalias{Simpson:2018cm} estimated $\nfe\sim3$). With the \gaia proper motion information and better photometry, we now conclude that the star is in fact a member of the cluster (see its placement on Figure \ref{fig:member_selection}).

\begin{figure*}
	\includegraphics[width=\textwidth]{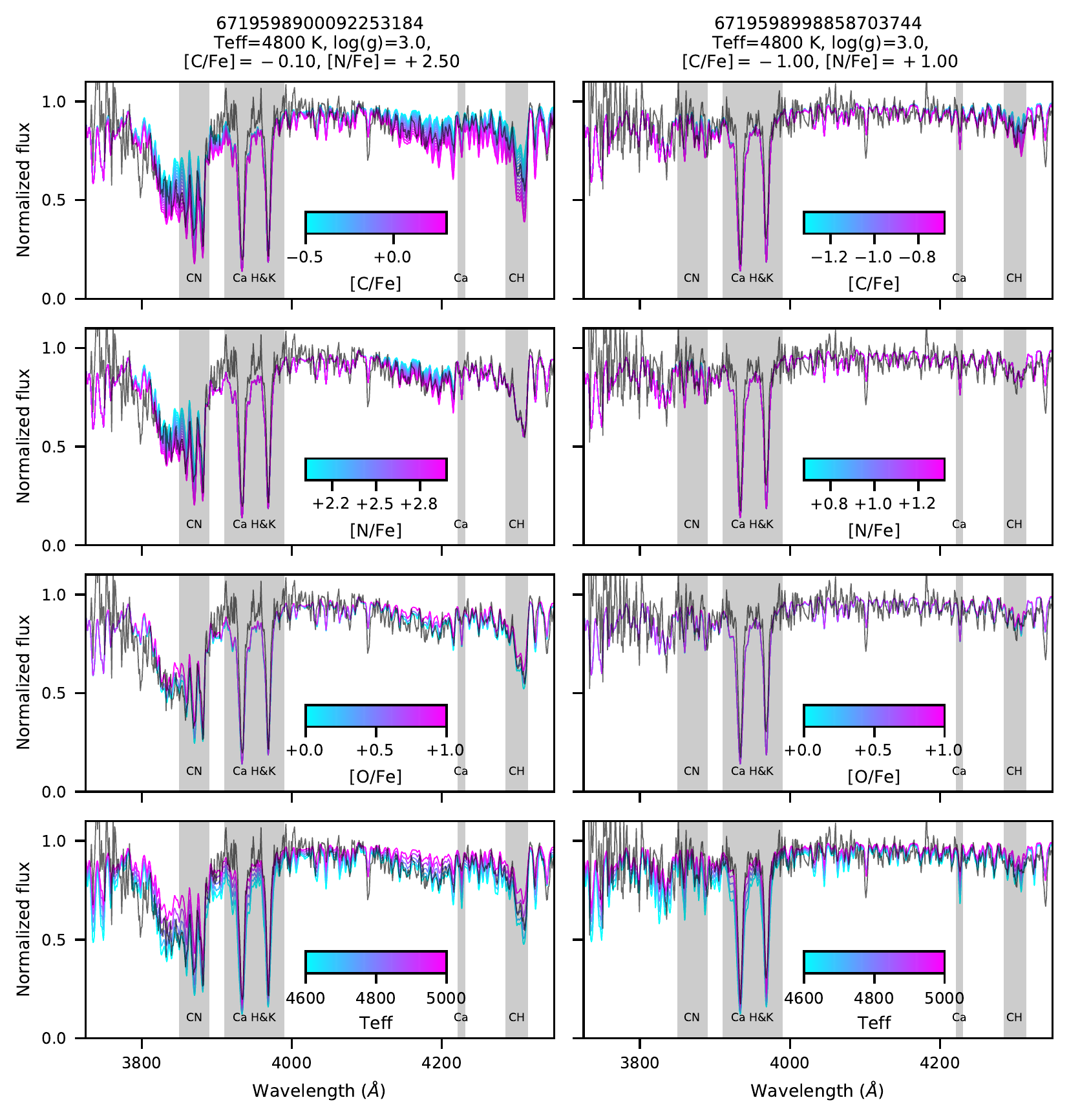}
    \caption{Comparison \response{of} synthetic spectra (blue through pink lines) to the observed spectra (black lines) for two \eso stars with \teff$\approx4800~$K. The left column ($\texttt{source\_id}=\cnstar$) is the CN-strong star, and the right column ($\texttt{source\_id}=6719598998858703744$) is a CN-normal star. In each row we show the effect of changing the abundance of a given parameter while keeping everything else constant: \cfe (first row), \nfe (second row), \ofe (third row), \teff (bottom row). Because of the \teff and \feh of these stars we find (as previous studies have shown) that there is little sensitivity in the CH and CN bands to \cfe and \nfe at the typical values found in metal-poor globular clusters.}
    \label{fig:observed_syn}
\end{figure*}

In this Section we infer the carbon and nitrogen abundances of \cnstarfull and six of the other bright RGB members of the cluster. For each star, we compared the observed spectrum to synthetic spectra created with \textsc{moog} \citep[][2017 Version]{Sneden1973}\footnote{The version used includes a proper treatment of scattering from \citet{Sobeck:2011dv} as implemented by Alexander Ji (\url{https://github.com/alexji/moog17scat})}. The model atmospheres are from ATLAS9 as created by \citet{Kirby:2011fm}. The atomic line list of neutral and singly-ionized species was created from Vienna Atomic Line Database \citep[VALD;][]{Piskunov1995,Ryabchikova1997,Ryabchikova2015,Kupka1999,Kupka2000} selecting all atomic lines between 3700 and 4600~\AA. This was supplemented with molecular line lists for $^{12,13}$C$^{14}$N \citep{Sneden:2014ki} and  $^{12,13}$CH \citep{Masseron2014}.

Due to the low resolution of the blue AAOmega spectra ($R\sim1200$), we cannot estimate the stellar parameters (i.e., \teff, \logg) directly from the spectra. Instead, we estimate the effective temperature using the $\textrm{J}-\textrm{K}_\textrm{S}$ infrared colour of the star and the empirical relationships from \citet{Alonso:1999go}, with the requirement that the stars had `A' quality photometry for their $\textrm{J}$ and $\textrm{K}_\textrm{S}$ magnitudes from 2MASS \citep{Skrutskie:2006hl}. This limits us to the brightest seven members of the cluster. We calculated the bolometric correction for the \gaia $G$ magnitude using \citet{Andrae2018}, and calculated the \logg assuming the stars have masses of $0.8 M_{\odot}$. We selected the closest atmosphere to within $\Delta\teff=100~K$ and $\Delta\logg=0.5$~dex.

With the atmospheric parameters set, we then estimate the carbon and nitrogen abundances. The carbon abundance was estimated by fitting the CH band at $\sim4300$~\AA\ by eye, and then similarly fitting the CN bands. This was iterated until a best fitting \cfe and \nfe was found. For \cnstarfull we estimate the errors to be $\pm0.2$~dex based upon the range of \cfe\ or \nfe that reasonably fits to the relatively noisy spectra. For the other stars, with weak-to-non-existent CN bands, the errors are much harder to quantify as a very large range of nitrogen abundances result in practically the same synthetic spectrum. The results of this analysis are shown in Table \ref{table:star_abund}, which shows that the abundances in the stars other than \cnstarfull are within the normal range for metal-poor globular cluster stars.

\begin{table}
\caption{Atmospheric stellar parameters and abundances for carbon and nitrogen for the seven brightest RGB members of \eso.}
\label{table:star_abund}
\begin{tabular}{lrrrr}
\hline
\texttt{source\_id} & \teff & \logg & \cfe  & \nfe \\
\hline
6719598075458648576 & $4200$ & $2.0$ & $-1.0$ & $0.5$ \\
6719599174970284928 & $4600$ & $2.5$ & $-0.8$ & $0.5$ \\
6719598998858703744 & $4800$ & $3.0$ & $-1.0$ & $1.0$ \\
6719598174224943104 & $4500$ & $2.5$ & $-0.8$ & $1.0$ \\
6719599170657398400 & $4400$ & $2.5$ & $-1.0$ & $1.0$ \\
6719598900092253184 & $4800$ & $3.0$ & $-0.1$ & $2.5$ \\
6719599101937916544 & $4600$ & $3.0$ & $-0.5$ & $1.0$ \\
\hline
\end{tabular}
\end{table}

Figure \ref{fig:observed_syn} shows the spectra of two \eso members: the CN-strong star \cnstarfull, and a CN-normal star with the same photometric \teff. Each column is for a particular star, with the continuum normalized observed spectrum repeated in each panel of the column. Each row shows the effect on the synthetic spectrum of changing a given parameter while holding the others constant: the top row column is changing \cfe by $\pm0.4$~dex about the best fitting \cfe value, the second row shows the same for \nfe, and the third row is $\ofe\in\{0.0,+0.5,+1.0\}$, and the bottom row is $\teff\in\{4600, 4800, 5000\}$~K.

We highlight the effect of \ofe because the oxygen abundances of these stars are unknown, and the oxygen abundance of a star can be important when considering carbon and nitrogen abundances derived from CH and CN molecular features. This is due to the molecular equilibria that exist in the stellar atmosphere between CH, CN, CO, and OH \citep[e.g.,][]{Russell1934}. In the third row of Figure \ref{fig:observed_syn} this manifests as a small effect in the synthetic spectra in the regions that are dominated by the carbon-including molecular features. Comparing the two extremes ($\ofe=0,1$), the strength of the CN and CH features are anti-correlated with the \ofe --- i.e., more oxygen in the atmosphere means more carbon is locked into CO and is not available to form CH and CN. The overall effect for these stars is relatively small, so we will assume a fixed value of $\ofe=+0.5$ for the rest of the analysis.

We also highlight the effect of \teff on the estimated abundances, which is somewhat complicated. The strengths of molecular features are strongly dependent on the surface temperature of the star, as shown in the bottom row of Figure \ref{fig:observed_syn}. If \cnstarfull is cooler than we have estimated (i.e., 4600~K instead of 4800~K), then the CH and CN features would be stronger, as the synthetic spectra illustrate. As such, less carbon is required in the stellar atmosphere to explain the strength of the CH features. But lowering the carbon abundance to fit the CH features would also result in the CN features being weaker. As such, the \nfe is relatively immune to small \teff changes, and we can be confident that this star is very nitrogen enhanced \response{--- making \cnstar cooler by 200~K requires an adjustment to the carbon abundance of $\Delta\cfe=-0.2$~dex, but no change to the nitrogen abundance. Similarly, there is little effect in changing the \logg: decreasing \logg by 0.5~dex only requires a $\Delta\cfe=-0.1$~dex adjustment}.

The CN-weak star (right column) shows that at this metallicity and temperature, there is little to no change in the spectrum when considering the range of carbon and nitrogen abundances typically found in globular cluster stars. Considering the right panel of the second row, varying the nitrogen through $+0.6<\nfe<+1.4$ barely registers any effect. It is only in the CN-strong star, where $\nfe\approx+2.5$, that abundance variations drive appreciable changes in the molecular line strengths. For \cnstarfull in addition to its large nitrogen abundance, it is relatively enhanced in carbon for a globular cluster star, with $\cfe=-0.1$, which can be seen in the relatively strong CH band at 4300~\AA, and also in hints of the C$_2$ bands redward of this.

\section{An NEMP star in \eso}\label{sec:discuss}

\begin{figure}
    \includegraphics[width=\columnwidth]{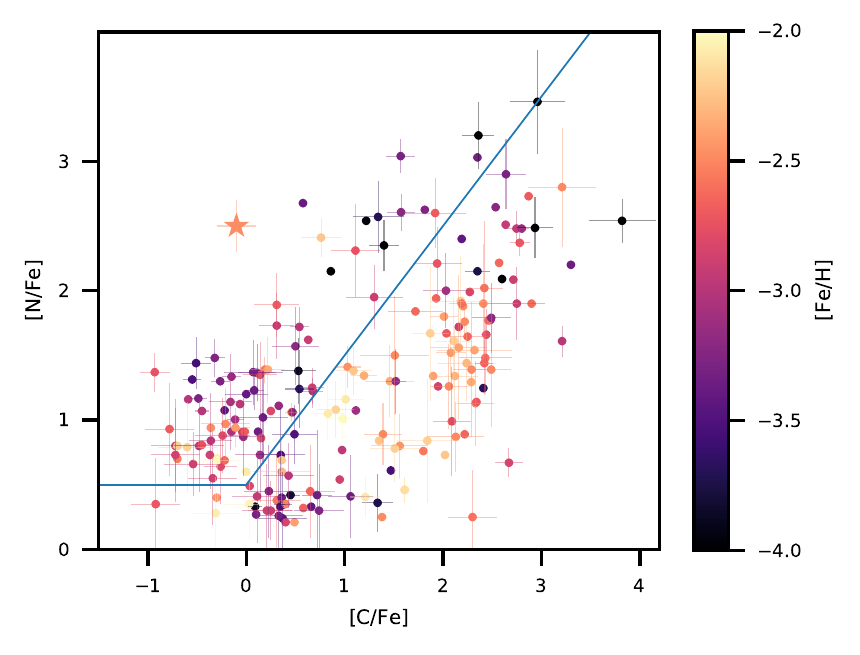}
    \caption{The \cfe and \nfe abundances of 199 metal-poor ($\feh<-2.0$) stars in the database of the Stellar Abundances for Galactic Archeology \citep[SAGA;][]{Suda2008} with $\nfe>0.2$. Each star is colour-coded by its literature \feh. Highlighted with a large star symbol is the CN-strong star \protect\cnstar. For stars with multiple measurements in SAGA, we have taken the mean of its values. The blue line shows the NEMP definition from \citet{Johnson2007} --- about 80 stars are above and to the left of the lines, indicating they are NEMP stars. If we require that stars have $\nfe>+2$, then there are only $\sim15$ stars.}
    \label{fig:n_enhanced}
\end{figure}

The star \cnstarfull has $\nfe=+2.5$, which is an extreme enhancement in the context of Galactic globular clusters. How do we interpret a star with such a high level of nitrogen enhancement? It is unlikely to be a result of the primordial enrichment and internal processing that normally influence light element abundances in globular clusters \citep[e.g.,][]{Charbonnel2007}. Those effects enhance nitrogen at the expense of carbon, and the carbon abundance in \cnstarfull is barely depleted at \cfe$=-0.1$. In addition, this level of nitrogen enhancement is a factor of 3 to 10 stronger than the maximum enhancement seen in other globular cluster giants.

Following the definition from \citet{Johnson2007}, this star is a nitrogen-enhanced metal-poor star as it has $\feh<-2$ with $\nfe>+0.5$ and $[\mathrm{N}/\mathrm{C}]>0.5$. In Figure \ref{fig:n_enhanced}, we show the 199 stars from the SAGA database with $\nfe>0.2$ and $\feh<-2$. In the \cfe--\nfe abundance plane \cnstarfull is somewhat isolated, with there being very few metal-poor stars that are both nitrogen enhanced and carbon poor. \response{There have been many searches for extremely metal-poor stars \cite[e.g.,][]{Caffau2013, Aguado2017, Starkenburg2017, Nordlander2019}, some with a bias toward or away from carbon enhancement, while NEMP stars have been less of a research focus. NEMP stars are also not as straightforward to identify observationally as CEMP stars.} The $3883~\hbox{\AA}$ CN band by which we initially identified this star is less prominent than the CH G band, and it is located at a shorter wavelength, where the \response{signal} in spectra of red giant stars is lower.


Mass transfer from a binary companion in the AGB phase is thought to be the source of the enrichment in NEMP stars \citep[e.g.,][]{Masseron2010, Hansen2016a}, because the anomalies in their abundance patterns resemble the result of nucleosynthesis in intermediate-mass AGB stars, including hot-bottom burning (HBB), the slow neutron capture process, and a high level of CNO cycle processing. \cnstarfull is not on the AGB itself (see its position in Figure \ref{fig:relative_cmds}), so the implication is that it must be in a post-mass transfer binary system. We have only one epoch\footnote{Although we have observed the cluster multiple times, this star only received one observation.} of low-resolution spectroscopy for this star, so we cannot make any strong statements about its binarity or its abundances of the s-process elements Ba or Sr. Additional spectroscopic observations of \cnstarfull, including radial velocity monitoring and at higher resolution, would allow us to test the AGB mass transfer scenario by establishing whether it is in a binary system and determining its full abundance pattern. We know that the CEMP-s stars, which have been proposed as post-AGB mass transfer binaries because of their enrichment in carbon and s-process neutron capture elements, have orbital periods and velocity semi-amplitudes of 30--3000 days and $<2$~\kms \citep{Hansen2015,Hansen2016,Hansen2016a}, indicating a likely search space for radial velocity variability for \cnstarfull.

Post-AGB mass transfer binaries certainly exist in globular clusters. CH stars (at lower metallicity) and barium stars (at higher metallicity), which are both understood as post-AGB mass transfer systems, are found in clusters as well as in the field \citep[e.g.,][]{McClure1984}. \cnstarfull is the only globular cluster star known with a nitrogen abundance high enough to require AGB mass transfer as an explanation. Its high nitrogen abundance indicates that the initial mass of its binary companion was at least 2.5--3~M$_{\odot}$, as lower-mass AGB stars produce mainly carbon and s-process elements \citep[e.g.,][]{Pols:2012bm, Karakas:2014jt}. The connection between AGB star mass and nucleosynthetic yields was also implicated in recent work by \citet{Fernandez-Trincado2019}, who identified a mildly metal-poor ($\feh=-1.08$) field giant with $\nfe=+0.69$ in APOGEE survey data. It has an excess abundance of the s-process element Ce, which indicates that its dynamically inferred binary companion was once a 5--7 solar mass AGB star.

The lack of NEMP stars in Galactic globular clusters can be at least partially explained by the fact that mass transfer from intermediate-mass AGB stars is required. In an environment with a limited number of stars, the number of high-mass stars that form is somewhat stochastic, even with an ordinary initial mass function. The distribution of binary mass ratios for 3$M_{\odot}$ stars is fairly flat \citep{Moe2016}, meaning that a value of 0.3, such as in this case, is not unexpected. Simulations with the BPASS models \citep{Eldridge2017} show that binary stars with masses of 3$M_{\odot}$ and 0.8$M_{\odot}$ can experience nitrogen-rich mass transfer.

\cnstarfull is currently unique among RGB stars in Galactic globular clusters. A considerable volume of spectroscopic data has been collected for giant stars in the very metal-poor clusters to investigate their nitrogen abundances (e.g., M15, M92, NGC~5466 by \citealt{Meszaros:2015fn}; M15, M92, NGC~5053 by \citealt{Smolinski2011}; NGC 6397 by \citealt{Pasquini2008} and \citealt{Carretta2005}), and no stars have previously been noted with such a dramatic enhancement in nitrogen abundance. Of course, there are RGB stars in these clusters that have not been observed spectroscopically in a way that would make an overabundance of nitrogen clearly visible, and the equally metal-poor clusters NGC 2419, M30, M68, NGC 4372, Palomar 15, NGC 6287, NGC 6426 and Terzan 8 do not have large catalogs of nitrogen abundance available. There may be additional NEMP stars waiting to be found in Galactic globular clusters.

\section*{Acknowledgements}
The data in this paper were based on observations obtained at the Anglo-Australian Telescope as part of programme S/2016A/13. We acknowledge the traditional owners of the land on which the AAT stands, the Gamilaraay people, and pay our respects to elders past and present. We are grateful to the then-AAO Director Warrick Couch for awarding us Director's Discretionary Time on the AAT which expanded our dataset.

\response{The authors thank the referee for their positive and helpful report, and} Amanda Karakas and JJ Eldridge for useful discussions. JDS and SLM acknowledge the support of the Australian Research Council through Discovery Project grant DP180101791. Parts of this research were conducted by the Australian Research Council Centre of Excellence for All Sky Astrophysics in 3 Dimensions (ASTRO 3D), through project number CE170100013. This work was written in part at the 2018 ASTRO-3D East Coast Writing Retreat.

This research includes computations using the computational cluster Katana supported by Research Technology Services at UNSW Sydney. The following software and programming languages made this research possible: \textsc{2dfdr} \cite[v6.46;][]{AAOSoftwareTeam:2015vz}, the 2dF Data Reduction software; \textsc{python} (v3.7.3); \textsc{astropy} \citep[version 3.2.1;][]{TheAstropyCollaboration:2013cd,TheAstropyCollaboration:2018ti}; \textsc{pandas} \citep[version 0.25.0;][]{McKinney:2010un}; \textsc{seaborn} (v0.9.0); Tool for OPerations on Catalogues And Tables \citep[\textsc{topcat}, v4.5;][]{Taylor:2005wx,Taylor:2006wv}; \textsc{matplotlib} \citep[v3.1.0][]{Hunter:ih,Caswell2019}. This work has made use of the VALD database, operated at Uppsala University, the Institute of Astronomy RAS in Moscow, and the University of Vienna. This work made use of v2.2.1 of the Binary Population and Spectral Synthesis (BPASS) models as described in \citet{Eldridge2017} and \citet{Stanway2018}.

This work has made use of data from the European Space Agency (ESA) mission {\it Gaia} (\url{https://www.cosmos.esa.int/gaia}), processed by the {\it Gaia} Data Processing and Analysis Consortium (DPAC, \url{https://www.cosmos.esa.int/web/gaia/dpac/consortium}). Funding for the DPAC has been provided by national institutions, in particular the institutions participating in the {\it Gaia} Multilateral Agreement.



%


\bsp    
\label{lastpage}
\end{document}